\documentclass[AMA,STIX1COL]{WileyNJD-v2}
\articletype{Article Type}%
\received{}
\revised{}
\accepted{}
\usepackage{graphicx}
\usepackage{epstopdf}
\usepackage{multirow,booktabs,xcolor}
\usepackage{balance,tabularx,subcaption,amsmath}
\usepackage{algorithm,algpseudocode}
\usepackage{mathtools,cuted,flushend}

\begin{document}

\title{SAR Analysis of Directive Antenna on Anatomically Real Breast Phantoms for Microwave Holography \protect\thanks{}}

\author[1]{Vineeta Kumari}

\author[2]{Gyanendra Sheoran*}

\author[3]{Tirupathiraju Kanumuri}

\authormark{AUTHOR ONE \textsc{et al}}

\address[1]{\orgdiv{Deprment of Electronics and Communication}, \orgname{National Institute of Technology}, \orgaddress{\state{Delhi}, \country{India}}}

\address[2]{\orgdiv{Department of Applied Sciences}, \orgname{National Institute of Technology}, \orgaddress{\state{Delhi}, \country{India}}}

\address[3]{\orgdiv{Department of Electronics and Electrical Engineering}, \orgname{National Institute of Technology}, \orgaddress{\state{Delhi}, \country{India}}}

\corres{*Gyanendra Sheoran. \email{gsheoran@nitdelhi.ac.in,sheoran.iitd@gmail.com}}

\presentaddress{}

\abstract[Summary]{Microwave holography utilizes antennae for detection of cancerous tissues in human breast. The safety assessment of an antenna is required prior to its radiation interaction with biological tissues. In this work, a specifically designed and developed Vivaldi antenna with directivity at $45^o$ for microwave holographic imaging has been presented. A detailed simulation  of Specific Absorption Rate (SAR) and thermal analysis due to the radiation exposure of the designed antenna on self designed anatomically real breast phantoms  is performed in Computer Simulation Technology (CST) environment. SAR values for 3D printed breast phantoms are also calculated experimentally. The quantitative results of SAR reveals the variation of $\pm$ 10\% in between the simulation and experimental results. It is found that the SAR due to the radiation of the designed antenna lies in the permissible limit. The thermal analysis in CST shows the rise of 0.096\% in temperature after the exposure of 2 hrs. Hence, the antenna can be used in microwave holography for breast cancer detection.}


%\jnlcitation{\cname{%
%\author{Williams K.}, 
%\author{B. Hoskins}, 
%\author{R. Lee}, 
%\author{G. Masato}, and 
%\author{T. Woollings}} (\cyear{2016}), 
%\ctitle{A regime analysis of Atlantic winter jet variability applied to evaluate HadGEM3-GC2}, \cjournal{Q.J.R. Meteorol. Soc.}, \cvol{2017;00:1--6}.}

\maketitle

\footnotetext{ }

\section{Introduction}
\label{Int}

Breast cancer is the most commonly diagnosed as non-melanoma cancer amongst women worldwide. Early detection is pivotal for the successful treatment of the disease. Microwave imaging is emerging as a promising substitution diagnostic tool, for breast cancer detection, with respect to already existing techniques. Its principle is based on the interaction of EM field (microwave) with living tissue (breast tissue) where it provides a contrast between the intrinsic dielectric properties of healthy and malignant tissue in the breast. This interaction is done with the antenna as it behaves as a source for transmitting and as a detector for receiving the EM field of the tissues \cite{r0}. Based upon the working principle of microwave holographic imaging, it is evident that the antenna is the back bone of the microwave holographic system. This interaction between the microwave and tissue could be short or long depending on the ability of antenna, power detector and processing unit. The antenna transmits microwave radiations at certain power which is expected to be lower or in line with the values of international standards \cite{r1}. At frequencies greater than 100 kHz, the primary cause of damage to the human tissue from EM exposure is due to the heating effects and non-electro-stimulation \cite{r2}. Hence, one must assess the exposure levels to avoid the tissue damage due to microwave radiations and ensure that interaction of antennas with human tissues are within the safety limits determined by the relevant authorities before utilizing them in any imaging system.

When the electromagnetic field is exposed on the human tissues, the energy associated with the electromagnetic wave is absorbed by the biological tissues \cite{r3}. The biological reaction under the exposure of electromagnetic field is of great significance while designing and developing a system. The extended face-to-face consequence of the electromagnetic field is the heating of the human tissue which may result in the increase of local temperature and hence affect the biological system. The heating effect on the tissue is normally quantified by the Specific Absorption Rate (SAR)\cite{r4}. 
The SAR value should not exceed beyond the level of exposure which becomes harmful as defined by the standards. The safety standards for human safeguard and protection from electromagnetic fields have been set by the International Commission on Non-Ionizing Radiation Protection (ICNIRP) and the Institute of Electrical and Electronic Engineers (IEEE). For instance, the maximum SAR described in the IEEE C 95.1:1999 guidelines are 1.6 W/kg for an average mass of 1g, whereas in the standard ICNIRP, is specified as 2 W/kg for an average mass of 10g. Whenever a near field system is designed for biomedical applications, the radiation exposure is a matter of concern. It is well documented in the literature \cite{r5} that the application of microwave imaging modalities represents a safe imaging technique.  However, these studies were based on  plane-waves illumination \cite{r2} in which power source was assumed to be of similar amplitude throughout to illuminate the whole breast tissues, which is not possible in real time as transmitter have decaying properties. Whereas, \cite{r6} used a simulated hemispherical breast model with homogeneous dielectric properties and \cite{r7} performed the simulation analysis with a lower power value i.e. -5 dBm. An analysis with  source power variation in both simulation and experiment should be needed for assessment of antenna exposure while designing an imaging system.

Generally, the SAR value depends upon different parameters such as the location of the antenna with respect to the human body, the strength of the radiation of the antenna, and the radiated power intensity \cite{r4}.
The proposed work focuses on the simulation of the SAR and thermal analysis of specifically designed and developed Vivaldi antenna. The SAR analysis on self fabricated 3D printed phantom for Vivaldi antenna is also performed experimentally, whereas, the thermal analysis is done in CST environment only. Authors have already proposed the utility of directive Vivaldi antennae in microwave holographic systems \cite{r9,r10} for imaging of metallic objects but its SAR analysis is required for safety assessment before its utilization in biomedical applications. To best of our knowledge, it the maiden attempt of employing vivaldi antenna and its SAR analysis for biomedical applications by microwave holography.

Rest of the paper is organised as : section 2 explains the design parameters of the  antenna used in the analysis, section 3 describes the results and discussion for the analysis of SAR absorption followed by conclusion in section 4.  
\section{Materials and Methods}
The methodology comprises of three steps: 
(1) Antenna Design  (2) Phantom Fabrication (3) SAR Analysis
\subsection{Antenna Design}
The simple structured specifically designed directive microstrip patch antenna i.e. Vivaldi antenna is developed to check its performance for the microwave holographic system. Furthermore, it will be used for early stage detection of cancer in breast tissues because it is lightweight, of low volume and low profile than the conventional antennas. This antenna is simulated using Computer Simulation Technology (CST) microwave studio software. The CST software is a supporting tool for 3D electromagnetic simulation that allows accurate result. 

\begin{figure}[htb!]
 \centering
\includegraphics[width=\linewidth]{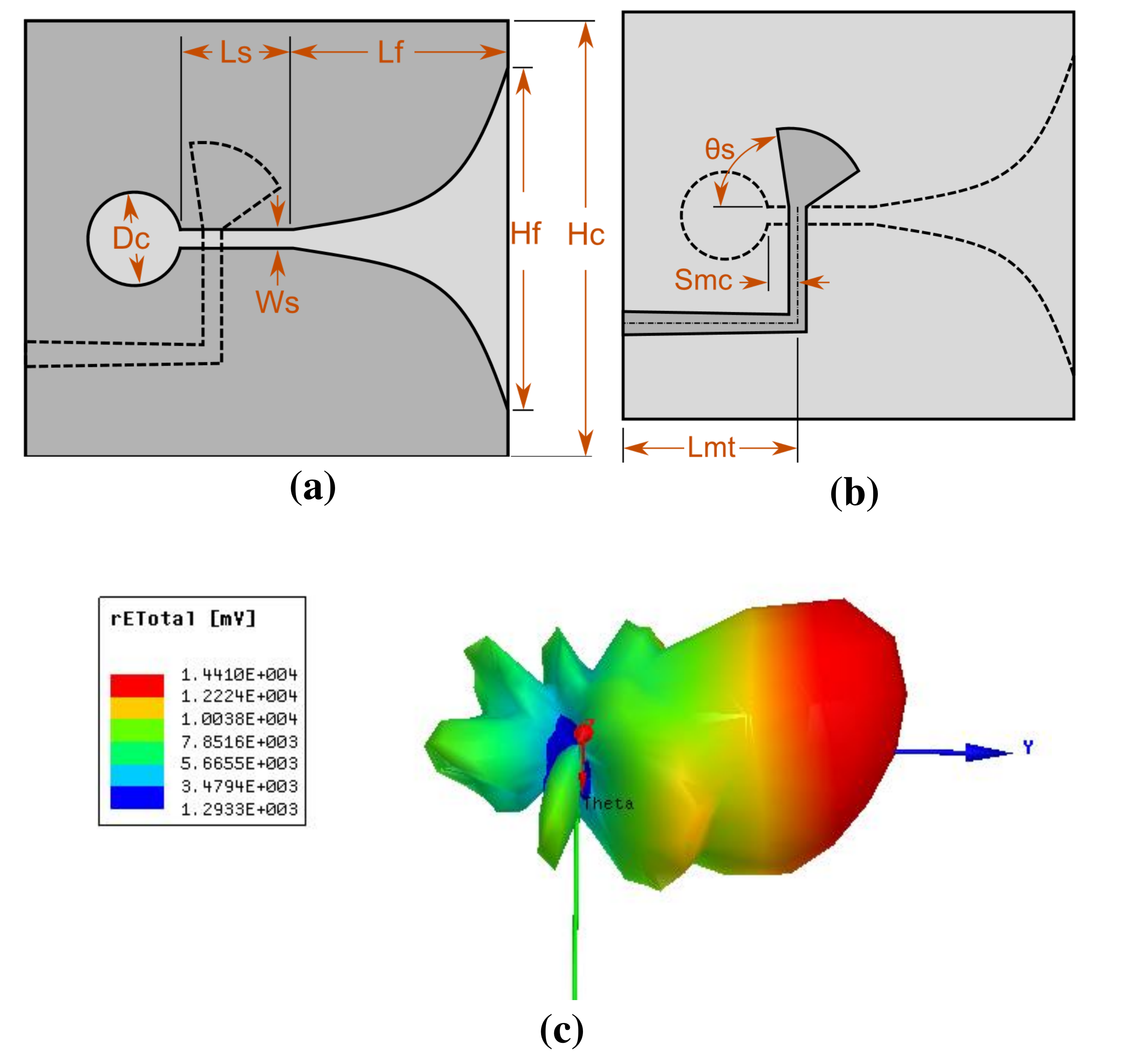}
\caption{Vivaldi antenna (a) front view (b) Rear view  (c) Gain of the antenna}
\label{f1}
\end{figure}

%\begin{table}[htb!]
%  \centering
%  \caption{Antenna Design Parameters for fig \ref{f1}(a) and (b)}
%    \begin{tabularx}{\linewidth}{>{\centering\arraybackslash}X>{\centering\arraybackslash}X}
%    \toprule
%    Antenna Design Parameters & Values (in mm) \\
%    \midrule
%    a  & 2.06 \\
%    \midrule
%    b     & 3.55 \\
%    \midrule
%    c     & 4.8 \\
%    \midrule
%    d     & 5 \\
%    \midrule
%    e     & 19.93 \\
%    \midrule
%    f     & 3.51 \\
%    \midrule
%    g     & 20.01 \\
%    \midrule
%    h     & 0.8 \\
%    \midrule
%    i     & 13.31 \\
%    \midrule
%    J     & 11.6 \\
%    \midrule
%    K     & 10.36 \\
%    \midrule
%    M     & 5.3 \\
%    \midrule
%    N     & 6.9 \\
%    \midrule
%    O     & 5.2 \\
%    \midrule
%    P     & 6 \\
%    \bottomrule
%    \end{tabularx}%
%  \label{tab:addlabel}%
%\end{table}%
The designed Vivaldi antenna is particularly developed and fabricated to work in the frequency band of 4-16 Ghz with the peak gain of 3.53 dB at 45 degrees. This antenna structure eliminates the design and fabrication of multiple antennae at different frequencies. In this structure, a metallized dielectric substrate is used with an exponentially tapered slot. A circular slot line cavity is connected to the narrow end of the flare. On the reverse side of the substrate the feeding is done and a microstrip line is flared to form a wide band radial quarter wave stub. The advantage of present feeding method is that both the antenna and its feed can be etched onto the same substrate. The base of the stub overlaps the slot line which is near to the circular cavity \cite{r11}. 
The simulated Ultra Wide Band (UWB) Microstrip-fed Vivaldi antenna and its designed structure is presented in Figs. \ref{f1}(a) and (b).
\begin{figure}[htb!]
 \centering
\includegraphics[width=\linewidth]{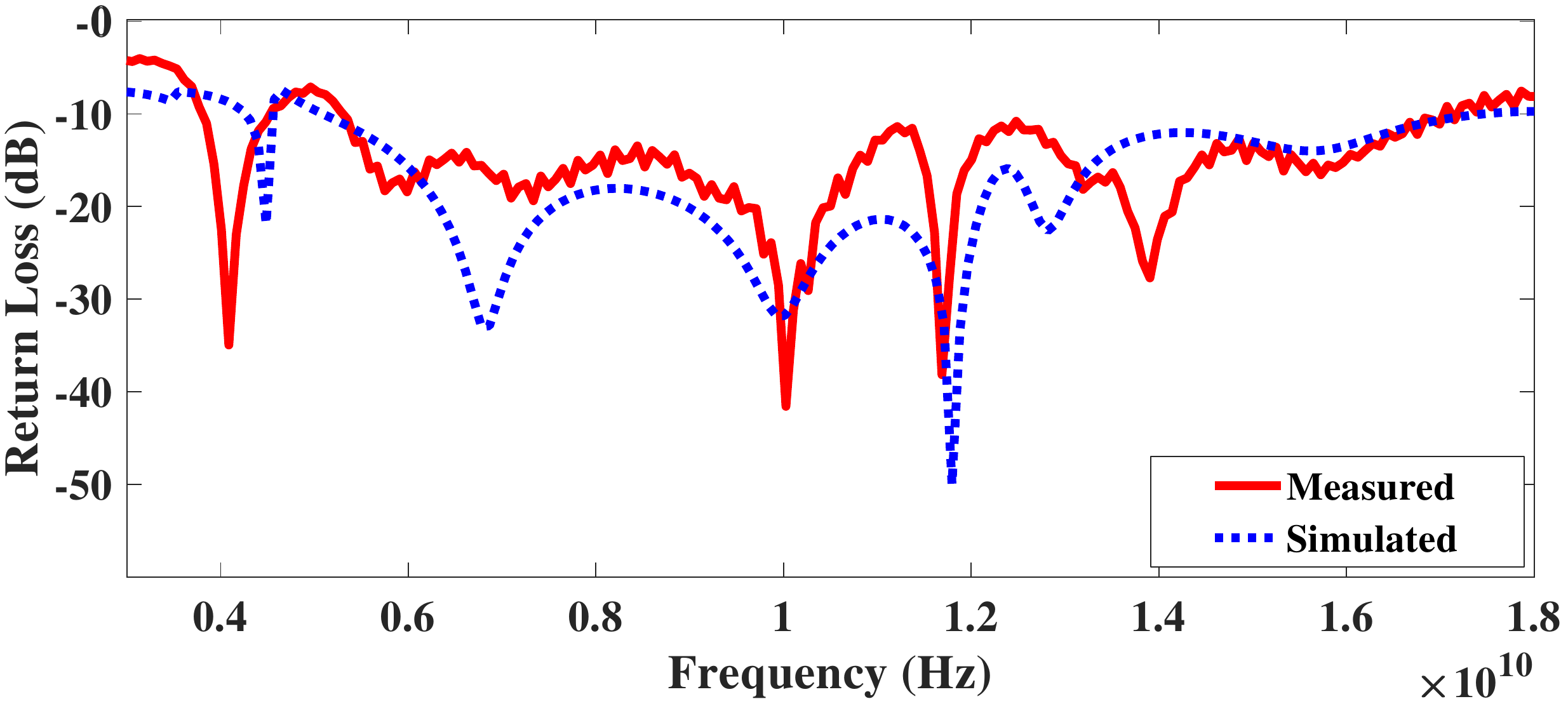}
\caption{ Simulated and measured return loss of Vivaldi antenna }
\label{f2}
\end{figure}
Fig. \ref{f1}(c) shows the performance of the antenna in terms of directive gain and radiation pattern. The design parameters are listed in the Table 1, Whereas, Fig. \ref{f2} shows the measured and simulated return loss values to show the operating frequency range of the antenna. 

\subsection{Breast Tissue Phantoms}
\subsubsection{Anatomically Real Phantoms}
The phantoms are designed and developed numerically using MR dataset of real patients (160 patients) with realistic dielectric properties mapped to the voxels. These voxel models depict the real dielectric properties of the patients of different ages at different frequencies ranging from 0.5 GHz - 20 GHz. The authors have provided the detailed composition used in the designing of the breast model in \cite{r8}. These models are simulated in MATLAB environment. The actual phantoms are simulated with a higher resolution i.e. 0.24 $\times$ 0.24 $\times$ 0.24 $mm^3$. But, due to computational constraints, the voxel resolution of the phantoms was taken as 1 $\times$ 1 $\times$ 1 $mm^3$ for the detailed analysis of microwave exposure.

\begin{figure}[htb!]
 \centering
\includegraphics[width=\linewidth]{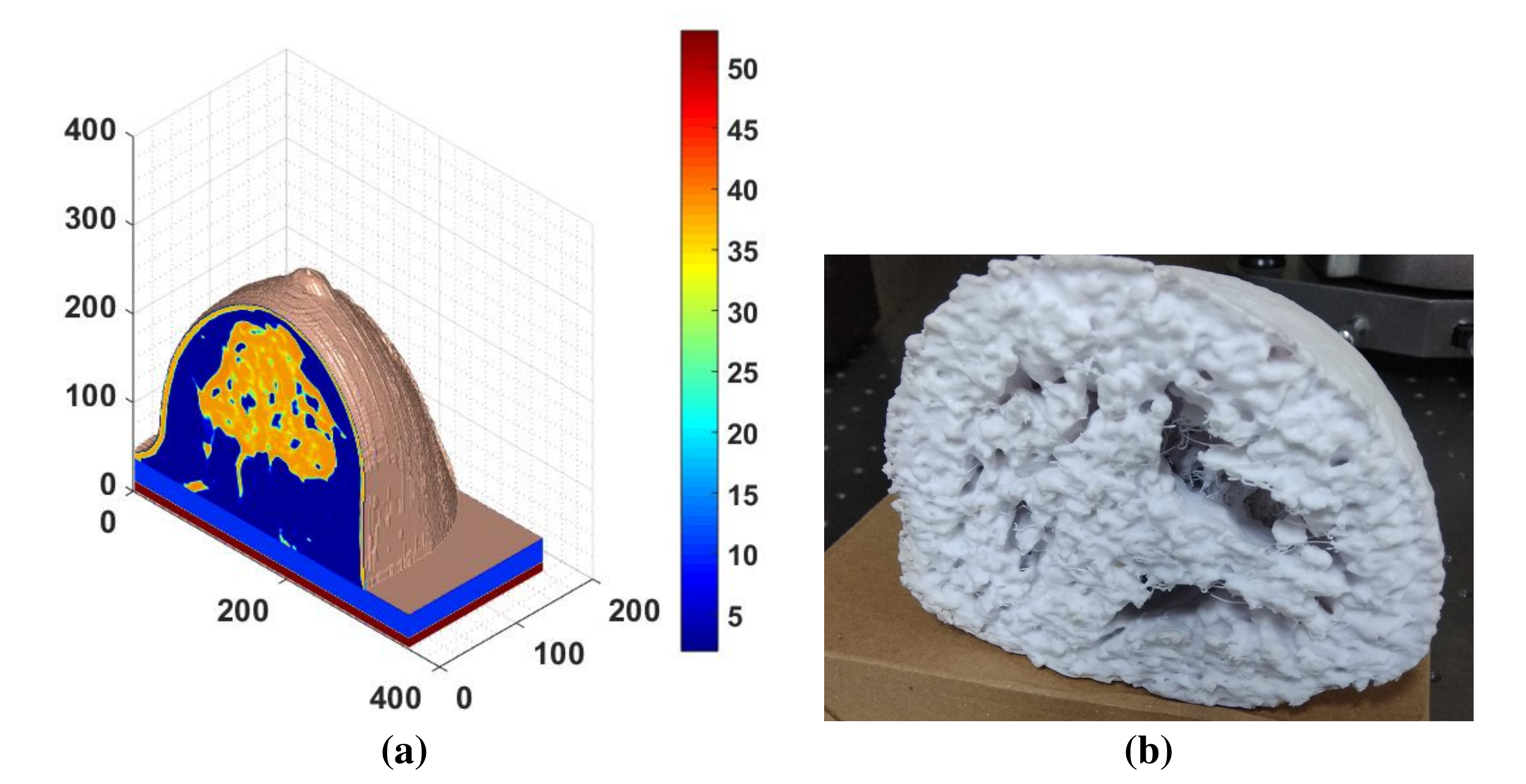}
\caption{ (a) Simulated and (b)fabricated 3D printed breast phantoms of ABS plastic (c) Voids filled with gelatine}
\label{f3}
\end{figure}

\begin{figure}[htb!]
 \centering
\includegraphics[width=\linewidth]{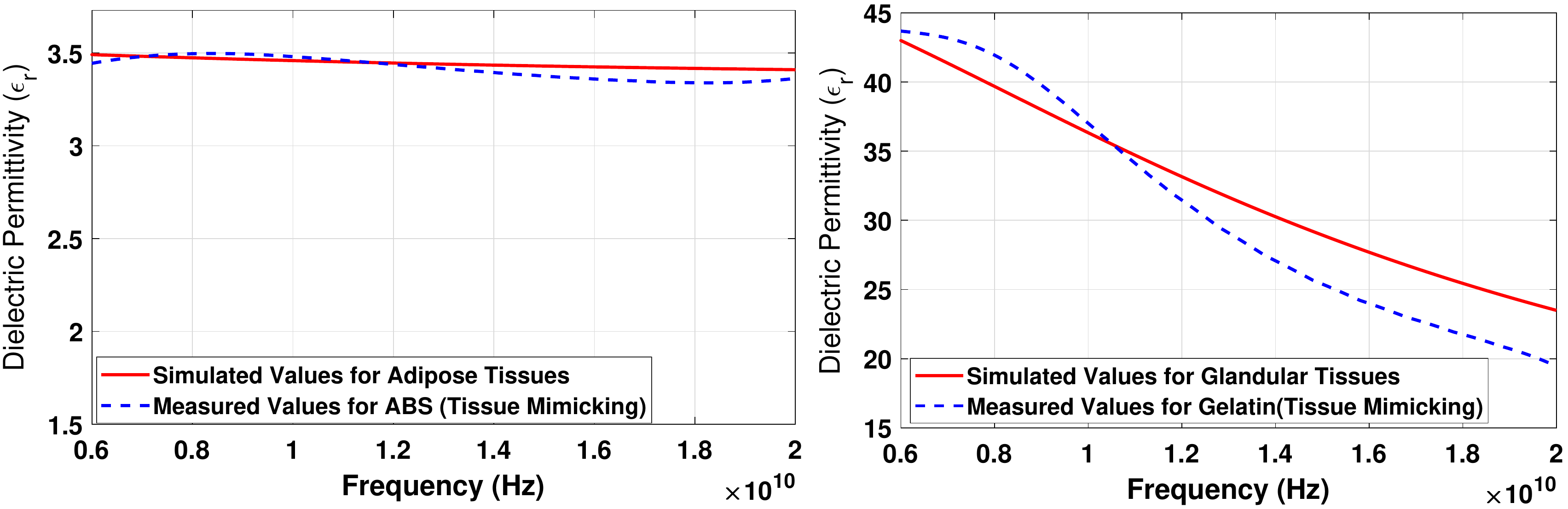}
\caption{ Simulated and measured dielectric properties}
\label{f4}
\end{figure}

In present work, the phantoms have been fabricated using tissue mimicking materials with similar properties of breast tissues.
The sagittal plane's view of the simulated phantom and 3D fabricated phantom of the same is shown in Fig. \ref{f3} (a) , (b) and (c) respectively.
Here, the MRI derived 3D printed tissue phantom is being used experimentally to demonstrate the microwave radiation absorption. The tissue portion of the breast phantom that is printed using Acronytrile Butadene Styrelene (ABS) plastic ($\epsilon_r=3.5$) corresponds to the locations of fat/adipose as shown in Fig. \ref{f3}. While, the voids in the model correspond to the fibroglandular tissues. The voids are filled with gelatin (20g) and  water (30g) mixture that constructs glandular tissue equivalent which provides a biologically relevant dielectric contrast with the adipose tissue. As, this 3D phantom is derived from the MR images of the real patients, it is correlated with the real distribution of breast tissues. 
The dielectric properties for the breast tissues are shown with the measured  dielectric properties of tissue mimicking materials (ABS and gelatine)in Fig. \ref{f4}. The measurements with frequencies ranging from 6-20 GHz were performed using VNA (Model No. ZNB40, Rohde and Schwartz) utilizing Nicolson-Ross-Weir (NRW) method \cite{r12,r13}. Authors have already simulated and presented the dielectric properties of adipose and fibroglandular tissues by employing the Debye/Cole-Cole model parameters \cite{r8}. The tissue mimicking material's dielectric properties are in accordance with the simulated values of tissues.
\subsubsection{CST Voxel Models}
The SAR analysis is also performed on the predefined voxel models of CST microwave studio. 
The CST voxel (available at:\url{http://www.cst.com/Content/Events/downloads/eugm2010           Bio Medical_RF_simulations_with_CST_MICROWAVE_STUDIO.pdf
}) family is a group of seven human model voxel data set created from seven different people with voxel resolution ranging from 0.98 $\times$ 0.98 $\times$ 10 $mm^3$ to 2.08 $\times$ 2.08 $\times$ 8 $mm^3$. We have selected only the breast tissues from these voxel models for a detailed analysis. 

\subsection{Specific Absorption Rate Analysis}
The rate of absorption of non-ionizing RF power per unit mass in human body tissues is termed as specific absorption rate \cite{r3}  (given by equation(1))
\begin{equation}
SAR = \,\sigma {\left| E \right|^2}/(\rho )
\end{equation}
Where, `E' denotes electric field(V/m), `$\sigma$' is the electrical conductivity (S/m), `$\rho$' gives the tissue density (kg/$m^3$). SAR values are measured in W/kg. Also, for radio frequency signals, SAR value is calculated for either 1 g (Australia, United States) or 10 g (Europe, Japan) of the simulated biological tissues. Where, partial (localized) non-occupational exposure is limited to a spatial peak value not exceeding 1.6 W/kg (Australia, United States) and 2.0 W/kg (Japan). The partial exposure of the SAR limit is recommended by the Council of the European Union and which is adopted by India given as 1.6 W/kg \cite{r3}.

\hspace{-0.75cm} The SAR analysis has been done in both ways: simulating the arrangement in CST environment utilizing Finite Difference Time Domain (FDTD) methodology and experimental investigation of the same with fabricated phantoms. The complete SAR analysis is done by importing the designed Vivaldi antenna to give exposure to the breast phantoms and CST voxel models in continuation of varying the distance between the antenna and breast models and power levels for a better insight of the analysis. For an optimal analysis of microwave exposure to the breast tissues, the voxel models for real patients are used. 

\section{Simulation and Experiment} 
\subsection{Simulation}
In the CST microwave studio, the antenna was located perpendicular to the model and the location of the antenna was further varied from 5 mm to 30 mm w.r.t. the fixed position of breast model for the analysis of exposure level. The breast models were exposed by the antenna at each above mentioned location with different power ranging from 1 mW to 0.5 W. 

To check the dependency of SAR on age, the analysis is done on CST simulator.
Age is an another factor which is responsible for the change in SAR values even if the other parameters involved in the experiment remain same. So, the analysis of the age wise exposure of microwaves is very necessary to determine the changes in SAR values with respect to breast density variations. The analysis is done on the basis of women's age to check the variation in their corresponding calculated SAR values. Since the breast density changes as the women's age grow \cite{r12}, therefore SAR values should change as per the change in the age of a woman.
The analysis are also performed with the designed antenna for analysing the effect of age on the SAR values. The voxel models of CST from age group (15 yrs - 80 yrs) are used to analyse the age wise SAR values. Although real numerical phantoms were also designed for real MRI patients \cite{r8} of each age group; but due to computational load of 160 patients we have used CST voxel models rather than the self-developed numerical phantoms for the particular age wise analysis. 

\subsection{Experimental Setup}
\begin{figure}[h!]
 \centering
\includegraphics[width=0.7\linewidth]{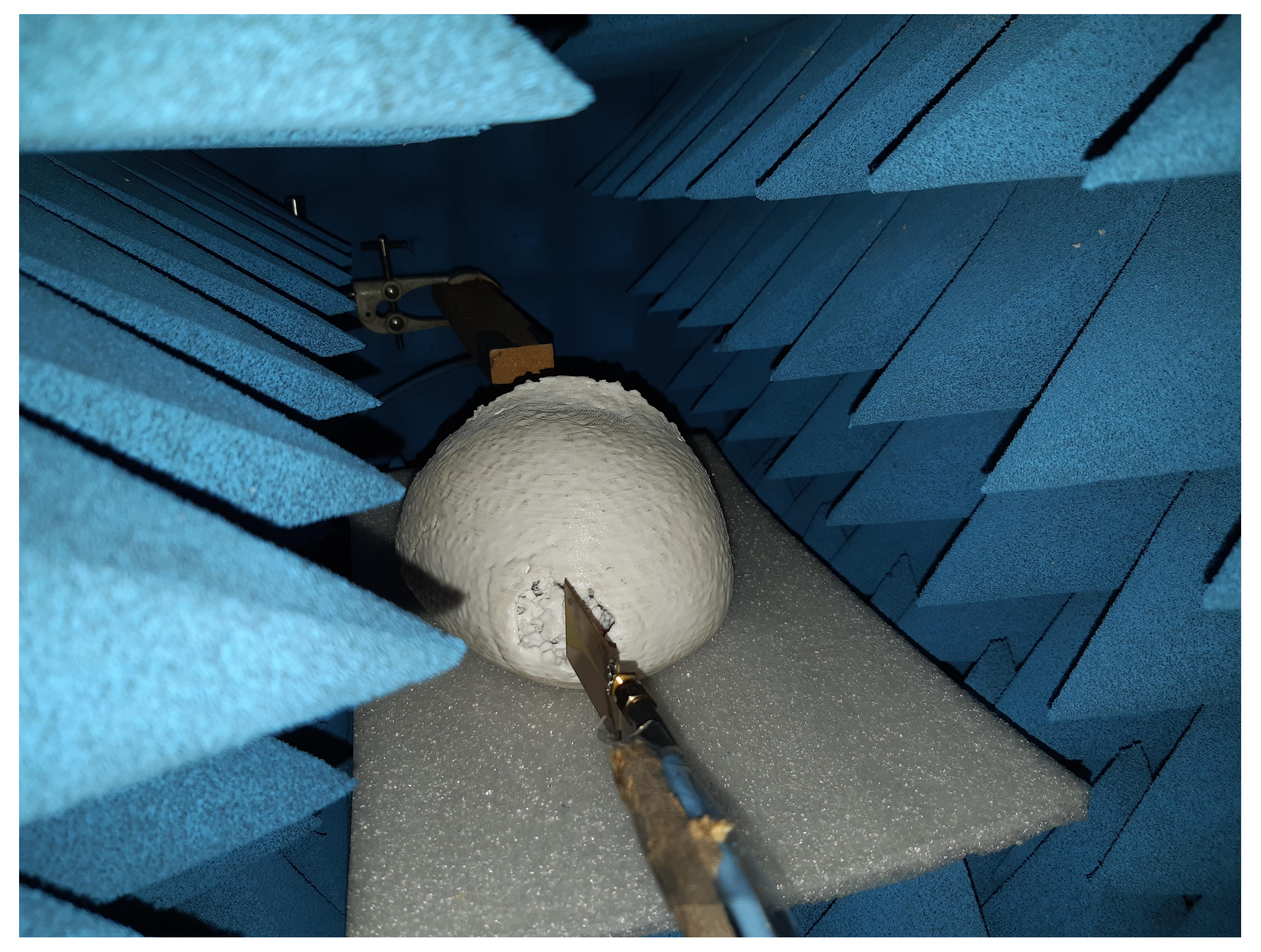}
\caption{ Experimental arrangement for SAR analysis with fabricated phantom}
\label{f5}
\end{figure}
To examine the specific absorption of the antenna radiations in real time, the fabricated Vivaldi antenna was placed in vicinity of the 3D printed breast phantom. The calibrations are done prior to the measurements using standard horn antennas which are connected to a thermal power sensor (R\&S NRP 33T) to perceive magnitude of the fed power. The microwave source (Rohde and Schwarz- Model No. SMB100A) was used for subjecting the breast phantom at different power levels ranging from 0dBm - 10dBm.
Thereafter, the measurements are done in which antenna under test are positioned consistent to the simulation. Fig.\ref{f5} shows the experimental arrangement for measuring the SAR values at a fixed distance of 5mm in between the in house fabricated phantom and antenna. The tissue mimicking phantom is prepared as explained in section 2.2. The whole set-up is stowed in microwave absorbers to avoid the external multiple reflections. This experiment is also performed without placing any phantom to subtract the background reflections.

\section{Results and Discussion}

\begin{figure}[htb!]
 \centering
\includegraphics[width=\linewidth]{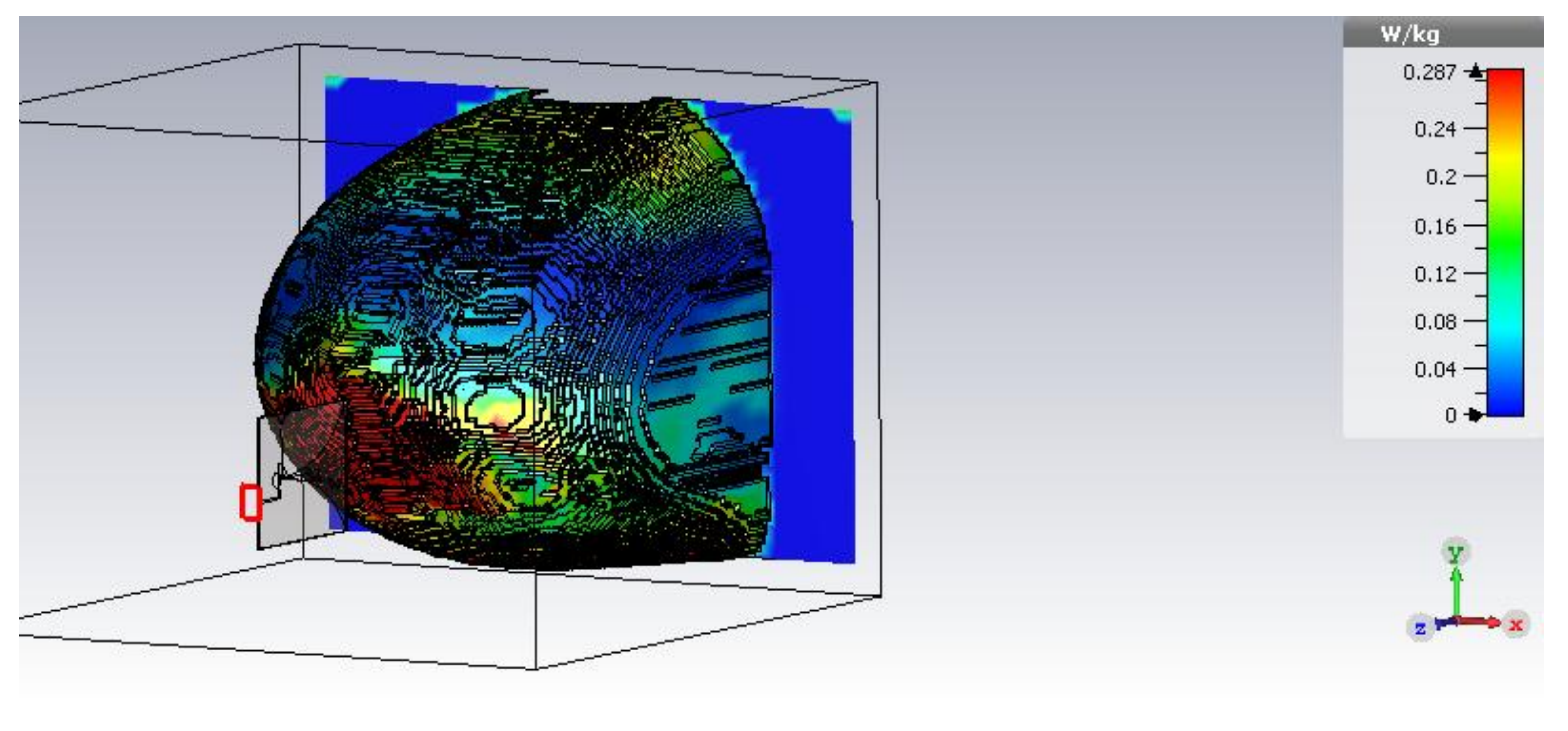}
\caption{Simulated SAR values with anatomical real phantoms in CST environment}
\label{f6}
\end{figure}

\begin{figure}[htb!]
 \centering
\includegraphics[width=\linewidth]{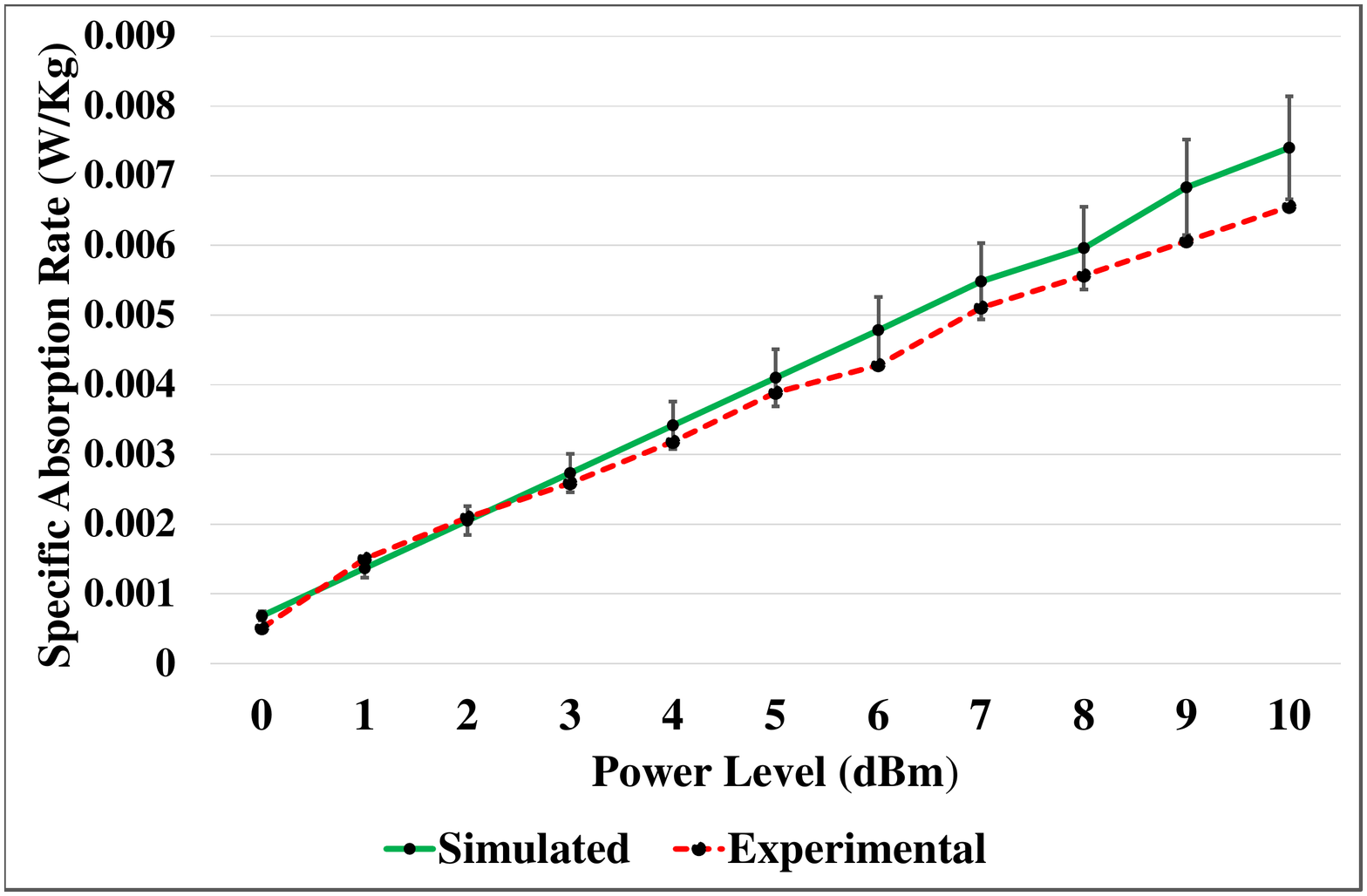}
\caption{SAR values for power level vs. distance variation in anatomically real phantom (a) 10g tissue (b) 1g tissue}
\label{f8}
\end{figure}

\begin{figure}[htb!]
 \centering
\includegraphics[width=\linewidth]{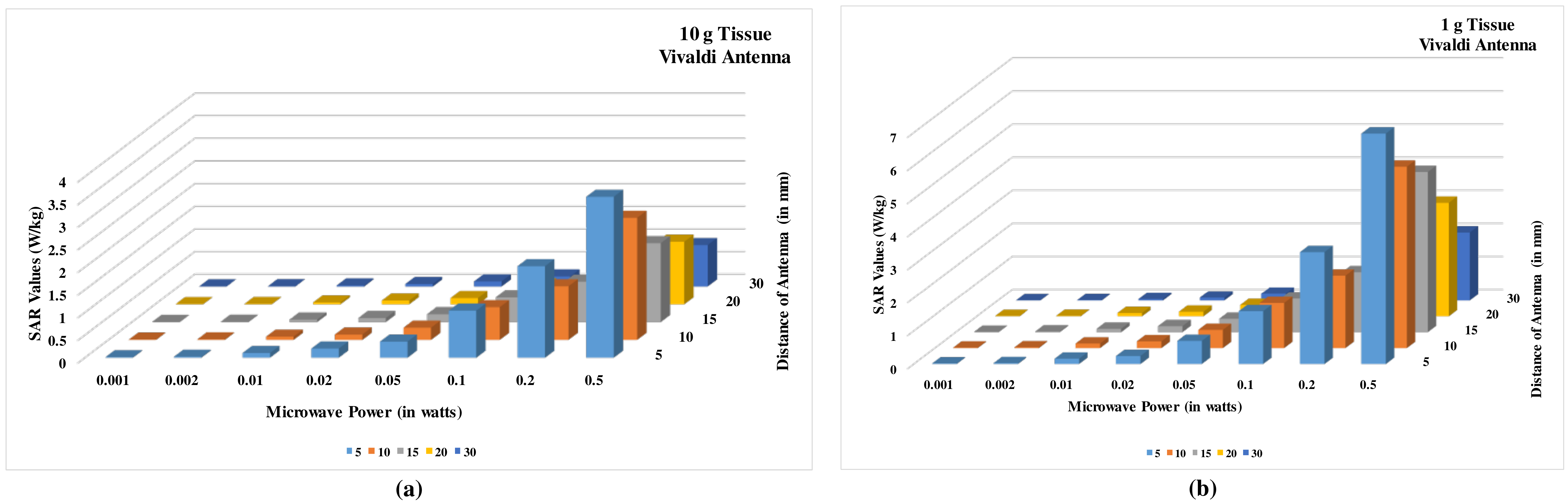}
\caption{SAR values for power level vs. distance variation for CST voxel model (a) 10g tissue (b) 1g tissue}
\label{f9}
\end{figure}

Fig. \ref{f6} shows the randomly simulated SAR values obtained after locating the Vivaldi antenna near to the anatomically real phantoms in the CST environment. 
Figs. \ref{f8} and \ref{f9} show the graphical form of the simulated  SAR values w.r.t the variation of microwave source power (1mW - 0.5W), and separation of antenna distance (5mm - 30mm) from the breast models.
Figure \ref{f8}(a, b) presents the SAR values for both the standards (1g tissue, 10 g tissue) by varying the separation in between the antenna and self developed tissue phantoms. The above figure depicts that the SAR value increases as the proximity of antenna with respect to the tissue increases. Also, SAR is directly proportional to the power of microwave source, which means, increase in source power level will also increase the absorption rate. The maximum SAR value is calculated and observed in the case of maximum power level(i.e. 0.5 W) and minimum separation distance (i.e. 5 mm) in between the tissue and an antenna. Similarly, figure \ref{f9}(a, b) shows the SAR values for both standards (1 g tissue, 10 g tissue) with the variation in the separation of an antenna distance and power levels for the antenna on the filtered out breast voxel models from the CST voxel family. 
The 3D plots in the figures are depicting that if the power value is set to 0.1 W or less, then the SAR will remain in the permissible limits irrespective of the antenna proximity. If the microwave power is increased beyond this limit then the SAR will also increase and crosses the permissible level for the near field locations of the antenna.

\begin{figure}[htb!]
 \centering
\includegraphics[width=\linewidth]{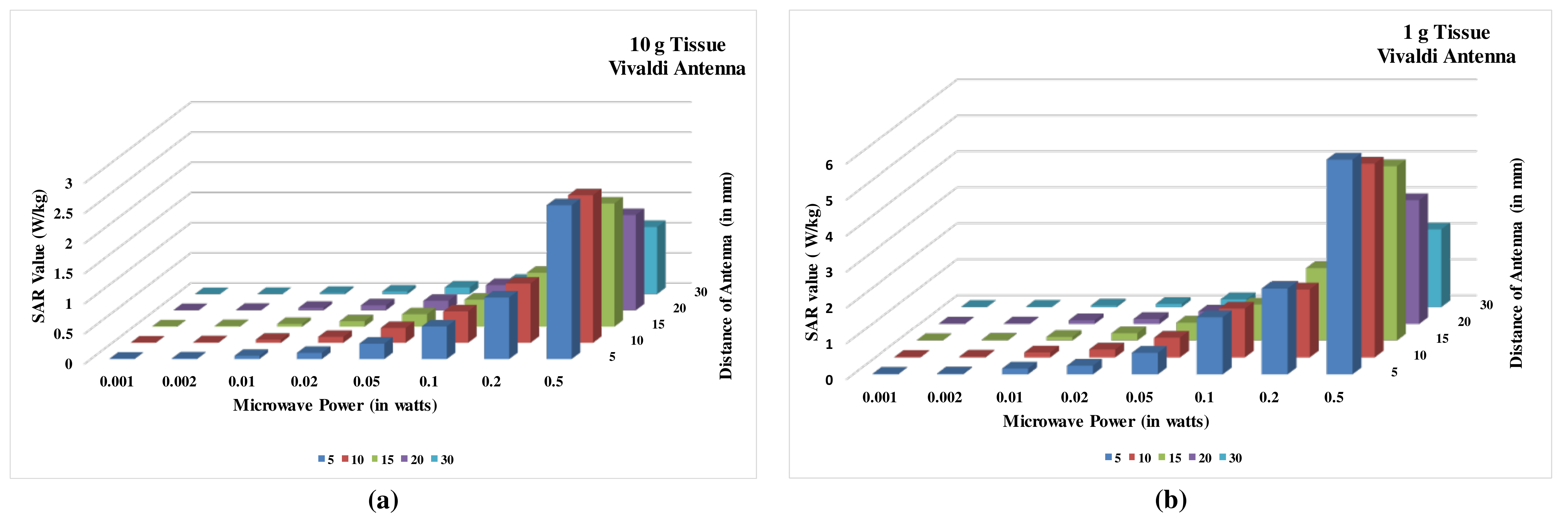}
\caption{ Simulated and experimental SAR values for (0dbm-10dbm) }
\label{f7}
\end{figure}
\begin{figure}[htb!]
 \centering
\includegraphics[width=\linewidth]{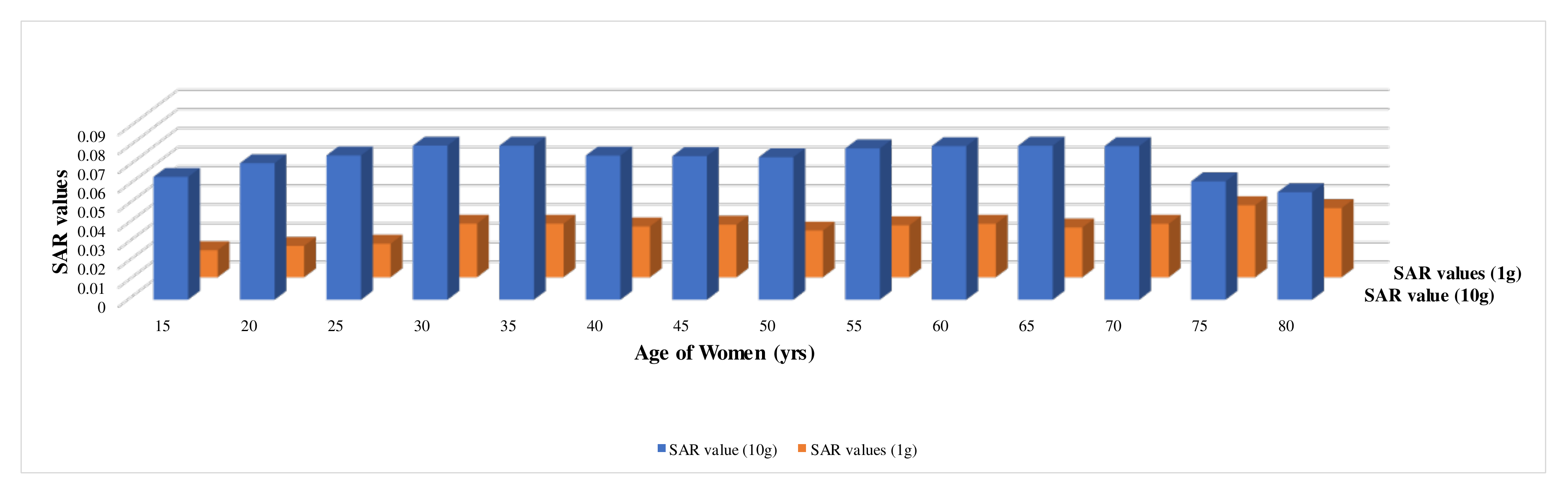}
\caption{Simulated SAR values for all age groups of women.}
\label{f10}
\end{figure}
The results presented in this section give a detailed analysis of the SAR values resulted under the exposure of the designed antenna. During the experiment, the power of the source was varied from 0dBm-10dBm. The SAR values for both simulation and experiments are shown in Fig.\ref{f7} for a fixed distance of 5mm. Here, the relative error is calculated and shown in the figure, which states that the experimental values are lying in the range of $\pm$ 10\% of the simulated values. The figure clearly supports the fact that the absorption rate for the designed antenna is within the permissible limits with all the power values. 

Fig.\ref{f10} shows the SAR analysis on the phantoms of different age group at randomly picked fixed distance and power corresponding to 5 mm and 1mw respectively. It is found that the change in SAR values is non-symmetrical for different patients because there is no fixed defined pattern of variation in the breast tissue density and age of the patients, breast density may differs for two individuals of same age \cite{r14}. The same is evident from the aforementioned figure, that the SAR values for different age groups vary independently, even though the distance of antenna and the microwave source power is fixed with the similar antenna placed in the vicinity of all breast models. 

Since, the study of temperature distribution in the human breast tissues is important to quantify any rise in the temperature due to the SAR values. Therefore, the experiment was performed in CST simulator for analysing the thermal exposure of the antenna. The tissues were exposed with the radiations for around 2 hours and the change in temperature was calculated by using the Pennes-Bioheat equation \cite{r15}. It was found that the temperature was varied from (310 K - 310.2 K) corresponding to 0dBm power and (310 K - 310.3 K) corresponding to 10dBm power. This supports the fact that there is no significant rise in the temperature while exposing the human breast tissues under the aforesaid power levels. 

\section{Conclusion}
In this work, the SAR is evaluated that arises due to the effect of self designed antenna's radiation on anatomically  real phantoms and CST voxel models. The detailed evaluation is done in CST simulator by considering the effecting parameters like antenna location, power variations, and age of patients. To compare and verify  the simulation results, experiments are also conducted on self fabricated tissue mimicking phantom. It is demonstrated that the fabricated phantom can be utilized for testing the microwave imaging systems in pre-clinical studies. The simulation analysis suggests that if the power level is decreased upto 10dBm or less, then the SAR values will remain in the permissible limits for all the locations of the antenna. The experimental study verifies and confirms that the designed antenna could be utilized in microwave imaging systems especially microwave holography for breast cancer detection.

%\backmatter

%\subsection*{Author contributions}
%
%This is an author contribution text. This is an author contribution text. This is an author contribution text. This is an author contribution text. This is an author contribution text. 

\subsection*{Financial disclosure}

None reported.

\subsection*{Conflict of interest}

The authors declare no potential conflict of interests.

\bibliographystyle{spmpsci} 

\clearpage
\begin{table}[htb!]
  \centering
  \caption{Antenna Design Parameters for fig \ref{f1}(a) and (b)}
    \begin{tabularx}{\linewidth}{>{\centering\arraybackslash}X>{\centering\arraybackslash}X}
    \toprule
    Antenna Design Parameters & Values (in mm) \\
    \midrule
    a  & 2.06 \\
    \midrule
    b     & 3.55 \\
    \midrule
    c     & 4.8 \\
    \midrule
    d     & 5 \\
    \midrule
    e     & 19.93 \\
    \midrule
    f     & 3.51 \\
    \midrule
    g     & 20.01 \\
    \midrule
    h     & 0.8 \\
    \midrule
    i     & 13.31 \\
    \midrule
    J     & 11.6 \\
    \midrule
    K     & 10.36 \\
    \midrule
    M     & 5.3 \\
    \midrule
    N     & 6.9 \\
    \midrule
    O     & 5.2 \\
    \midrule
    P     & 6 \\
    \bottomrule
    \end{tabularx}%
  \label{tab:addlabel}%
\end{table}%
\listoffigures
\listoftables
\end{document}